\documentstyle[11pt,ncs,a4]{article}
\include{psfig}
 
\def \square {\hbox{$\sqcup\!\!\!\!\sqcap$}} 

\def\sp#1{{}^{#1}}                              
\def\sb#1{{}_{#1}}                              

\def\bra#1{\left\langle #1\right|}             
\def\ket#1{\left| #1\right\rangle}             
\def\vev#1{\left\langle #1\right\rangle}       

\newcommand{\be}{\begin{equation}}
\newcommand{\ee}{\end{equation}} 
\newcommand{\bea}{\begin{eqnarray}}
\newcommand{\eea}{\end{eqnarray}}

\catcode`\@=11 
\@addtoreset{equation}{section}
 
\catcode`\@=11

\begin{document}

\begin{titlepage}

\begin{flushright} 
{\tt 	 FTUV-96-79\\IFIC/96-88}
 \end{flushright}

\bigskip

\begin{center}

{\bf{\LARGE A Quantum Model of Schwarzschild \\
Black Hole Evaporation}}\footnote{Work partially supported by the 
{\it Comisi\'on Interministerial de Ciencia y Tecnolog\'ia 
and  DGICYT}.}

\bigskip 
J. Cruz$^1$\footnote{\sc cruz@lie.uv.es},
A.  Mikovi\'c$^{1,2}$\footnote{\sc mikovic@castor.phy.bg.ac.yu} and
 J. Navarro-Salas$^1$\footnote{\sc jnavarro@lie.uv.es}
 \end{center}


\footnotesize

\begin{enumerate}	                 
\item Departamento de F\'{\i}sica Te\'orica and 
	IFIC, Centro Mixto Universidad de Valencia-CSIC.
	Facultad de F\'{\i}sica, Universidad de Valencia,	
        Burjassot-46100, Valencia, Spain. 
\item Institute of Physics, P.O. Box 57, 11001 Belgrade, Yugoslavia. 
\end{enumerate}
\normalsize 

\bigskip

\begin{abstract}			

We construct a one-loop effective metric
describing the evaporation phase of a Schwarzschild black hole 
in a spherically symmetric null-dust model. This is achieved by 
quantising the Vaidya solution and by chosing a time dependent quantum
state. This state describes a black hole which is initially in thermal 
equilibrium and then the equilibrium is switched off, so that the
black hole starts to evaporate, shrinking to a zero radius in a finite
proper time. The naked singularity appears, and the Hawking flux diverges at
the end-point. However, a static metric can be imposed in the future of the 
end-point. Although this end-state metric cannot be determined within our 
construction, we show that it cannot be a flat metric.


\end{abstract}



\end{titlepage}

\newpage

\section{Introduction}

The two-dimensional dilaton gravity models turned out to be very useful
toy models of black hole formation and evaporation \cite{Strom}. Their relevance
for 4d black holes comes from the fact that the spherically symmetric scalar
field collapse can be described by a 2d dilaton gravity action
\be
S={1\over2}\int dr dt\sqrt{-g}e^{-2\phi}\left[\left(R+2\left(\nabla
\phi\right)^2 + 2e^{2\phi}\right)-{1\over2}
G\sum_{i=1}^N\left(\nabla f_i\right)^2\right]\>,\label{11}
\ee
where $G$ is the Newton constant and the 4d line element $ds_4$ is related to 
the 2d line element $ds$ by
\be
ds\sb 4\sp 2 =ds^2 +  e^{-2\phi} d\Omega^2\>.\label{12}
\ee
$R$ is the 2d scalar curvature associated with the 2d metric $g_{\mu\nu}$,
$\phi$ is the dilaton field
and $f_i$ are $N$ matter scalar fields. 
These fields depend only on time $t$ and radial coordinate $r$, while the
angular dependence resides in $d\Omega^2 $.
The spherically symmetric collapse was studied by several authors \cite{BUH}, 
and the problem of determining a semiclassical metric which
includes the back-reaction of the Hawking radiation 
is still unsolved. This is related to the fact that the classical
equations of motion are not solvable. In contrast to this,
a string theory inspired 2d dilaton gravity model \cite{CGHS}
\be
S_0={1\over2}\int d^2x\sqrt{-g}\left[e^{-2\phi}\left(R+
4\left(\nabla\phi\right)^2+4\lambda^2\right)-{1\over2}
\sum_{i=1}^N\left(\nabla f_i\right)^2\right]\>,\label{13}
\ee
is classicaly solvable, and its solution describes a formation of a 2d 
black hole. The quantization of (\ref{13}) is made simpler by the fact that 
the matter fields propagate freely 
\cite{Mik92,Mik93,SVV,HKS,Mik95,Mik96,Jackiw,Kuchar}, 
so that the one-loop \cite{RST, BPP, Mik95},
and the two-loop \cite{MR} effective metrics were obtained. Therefore one can
study analytically the back-reaction effects in this 2d model. 

In this paper we are
going to study a more realistic 2d dilaton gravity model, which will have 
some of 
the nice features of (\ref{13}) but it is going to describe a 4d black hole. 
We will study 
\be
S={1\over2}\int d^2x\sqrt{-g}\left[e^{-2\phi}\left(R+2\left(\nabla
\phi\right)^2 + 2e^{2\phi}\right)-{G\over2}
\sum_{i=1}^N\left(\nabla f_i\right)^2\right]\>,\label{14}
\ee
whose 4d interpretation is that of a self-gravitating
spherically symmetric null-dust cloud. When compared to the action (\ref{11}),
one notices that a simplifying feature of (\ref{14}) is that the matter fields 
do not couple to the dilaton, so that one obtains  
free-field matter equations of motion
in the conformal gauge, in analogy to the action (\ref{13}). 
This means that the task of determining the back-reaction in the
model (\ref{14}) is going to be simpler than in the model
(\ref{11}). Still, the quantization of (\ref{14}) is complicated
by the fact that the general solution of the 
equations of motion is not known. However, in a special case when 
\be
ds^2=-\left(1-{2m\left(v\right)\over r}\right)dv^2+2dvdr\>,\label{15}
\ee
where $\exp(-\phi)= r$, $G=1$, the classical
solution for $f_i =f_i (v)$ is given by
\be
{dm (v)\over dv} = T_{vv} (v)
= {1\over2}\sum_{i=1}^N\left({df\sb i\over dv}\right)^2 \quad,\label{16}
\ee
which is the well-known Vaidya solution \cite{Vaidya}. It
describes a collapse of
a spherically symmetric null-dust cloud. The equations (\ref{15}) 
and (\ref{16}) will be the starting
point for our quantization procedure, from which we will determine an
effective metric describing the one-loop back-reaction effects. 
Note that the back-reaction effects in the model (\ref{14}) have been 
studied in \cite{Lowe, Parentani} where the one-loop back-reaction
has been modeled by adding a Polyakov-Liouville
term to the action (\ref{14}). However, the resulting equations of motion 
are not solvable,
and only a numerical study has been done. 

In this paper we will perform
an operator quantization of the equations (\ref{15}) and (\ref{16}), 
so that an explicit expression for a
one-loop effective metric will be obtained. This metric will describe the
evaporation of a Schwarzschild black hole which was initially in a 
thermal equilibrium state. This is achieved by using a
quantization formalism developed in \cite{Mik95,Mik96}, and 
by using the idea of thermal bath removal \cite{CN},
which was developed in the case of the 2d model (\ref{13}). 
We first show that the idea of 
thermal bath removal
can be naturally formulated in the operator formalism, where it corresponds
to the introduction of a time dependence in the 
Heisenberg quantum state of the system.
This time dependence can be attributed to the external forces which switch off
the thermal equilibrium. Then we implement this idea to the model
(\ref{14}) and obtain a one-loop metric whose properties we study.

\section{Operator formalism and thermal bath removal in the CGHS model}

The general solution of the classical equations of motion following from 
(\ref{12})
in the conformal gauge $ds^2=-e^{2\rho}dx^+dx^-$ 
are (up to constant shifts in the $x^{\pm}$ coordinates)
\be
e^{-2\rho}=e^{-2\phi}=-\lambda^2 x^+x^- - F_+ - F_- + {M\over \lambda}\>,
\label{21}
\ee
\be
f_i = f_{i+}\left(x^+\right) + f_{i-}\left(x^-\right)\>,\label{22}
\ee
where
\be
F_{\pm}=\int^{x^{\pm}}\int^{x^{\pm}} T_{\pm\pm} (x^+ ) \>,
\label{23}
\ee
and $T_{\pm\pm} = \frac12 (\partial_{\pm}f)\sp 2$.
$M/\lambda$ is an integration constant, and the residual conformal 
invariance has been fixed by the so-called
"Kruskal" gauge $\rho=\phi$.

From the expressions (\ref{21}), (\ref{22}) and (\ref{23}) 
it is clear that the independent degrees of 
freedom are those of the matter fields.
Therefore a reduced phase space (RPS) quantization should give a physical 
Hilbert space which
coincides with the Hilbert space of massless scalar fields propagating on 
a flat background \cite{Mik93,Mik95}. As far as the problem of diffeomorphism 
anomalies is concerned, it is formally avoided in the RPS 
quantization, although it may 
be hidden in the non-covariant form of the gauge-fixed theory. However, 
an anomaly-free Dirac quantization of the CGHS model gives the same
physical Hilbert space as the RPS quantization
\cite{Kuchar}, which gurantees the diffeomorphism invariance of the RPS results.
The dynamics is generated by the free-field hamiltonian of N massless scalar 
fields, and therefore the quantum evolution is unitary. 
The effective metric is determined by
$ ds\sp 2 = -\vev{e\sp{2\rho}} dx\sp + dx\sp - $. 
The effective conformal factor
can be evaluated perturbatively by using a matter-loop expansion
\cite{Mik95,Mik96}, so that at the one-loop order one obtains
\be
e\sp{-2\rho\sb 1} = \bra{\psi_0}e\sp{-2\rho}\ket{\psi_0}=
-\lambda^2 x^+x^- - \vev{F_+}- \vev{F_-} + {M\over\lambda} \>,\label{24}
\ee
where now $F_\pm$ are operator valued expressions (\ref{23}) in the 
Heisenberg picture.
The initial state $\ket{\psi_0}$ can be chosen to be a coherent state
$e\sp{A_+} \ket{0_{\sigma\sp +} } \otimes \ket{0_{\sigma\sp - }} $,
corresponding to a left-moving pulse of matter, where 
$\sigma^{\pm}$ are asymptotically flat dilaton vacuum coordinates
$\left(\lambda x^{\pm}=\pm e^{\pm\lambda\sigma^{\pm}}\right)$ and
$\ket{0_{\sigma\sp + }}\otimes\ket{0_{\sigma\sp -}}$ is the corresponding 
vacuum. If the
normal ordering in $T_\pm$ is chosen to be with respect to the Kruskal
vacuum $\ket{0_{x\sp +}}\otimes\ket{0_{x\sp -}}$, then \cite{Mik96}
\be
\bra{\psi_0}T_{++}\ket{\psi_0}=-{N\over 48\pi\left(x^+\right)^2}+
{1\over2}\left(\partial_+f\right)^2 \quad,\quad
\bra{\psi_0}T_{--}\ket{\psi_0}=-{N\over48\pi\left(x^+\right)^2}\>,\label{27}
\ee
where we have chosen the conventional normalization of the flux \cite{BD}, 
so that there is no a factor of $1/\pi$ in (\ref{13}).
The expression (\ref{24}) then gives an evaporating black hole solution 
corresponding to the one-loop effective action of \cite{BPP}
\be
S=S_0-{N\over96\pi}\int d^2x\sqrt{-g}R\square^{-1}R
-{N\over24\pi}\int d^2x\sqrt{-g}\left(R\phi-\left(\nabla\phi
\right)^2\right)\quad.\label{28}
\ee

Note that one can consider
a different process of black hole evaporation, if a different
initial state is chosen.
Instead of the boundary conditions (\ref{27}), which 
correspond to the gravitational collapse situation,
one can consider an evaporation process where initially one has a black hole
in thermal equilibrium and at $x^+=x^+_0$ 
the incoming thermal flux is switched off \cite{CN}.
This process can be described by the boundary conditions
\be
\vev{T_{++}}=-{N\over48\pi\left(x^+\right)^2}\theta\left(x^+-x^+_0\right)\quad,
\quad
\vev{T_{--}}=0\quad.\label{210}
\ee
It is not difficult to see that the boundary conditions 
(\ref{210}) correspond to the following state $\ket{\Psi_0}$
\be \ket{\Psi_0} = \theta (x\sb 0\sp + - x\sp + ) \ket{0\sb{x\sp +}}
\otimes \ket{0\sb{x\sp -}}  + 
\theta (x\sp + - x\sb 0\sp + )\ket{0\sb{\sigma\sp +} }\otimes 
\ket{0\sb{x\sp -}} 
\,. \label{211} \ee
A novel feature of (\ref{211}) is that the Heisenberg state $\ket{\Psi_0}$ 
now depends on time, which reflects the
nature of the new process where an external force has to be used in order to
do the switching. It follows from (\ref{211})  that
\be 
\vev{T\sb{\sigma\sp + \sigma\sp +}}={N\lambda^2\over48\pi}
\theta ( x\sb 0\sp + - x\sp + )
\quad,\quad \vev{T\sb{\sigma\sp - \sigma\sp -}} = {N\lambda^2\over48\pi} \quad,
\label{212} 
\ee
where the normal ordering in $T_{\sigma\sp{\pm} \sigma\sp{\pm}}$ is with 
respect
to the dilaton vacuum, so that the incoming constant thermal 
flux has been switched off.  

The one-loop solution for $x^+ > x_0^+$ is now given by
\be
e^{-2\phi}=-\lambda^2 x^+\left(x^-+\Delta\right)-{N\over48\pi}\log\left(
{x^+\over x_0^+}+1\right)+{M\over\lambda}\>,\label{213}
\ee
where $\Delta=-N/48\pi\lambda^2 x_0^+$.
The asymptotically flat coordinates $\tilde\sigma^{\pm}$ at the future 
null-infinity are given by
\be
\tilde\sigma^+ = \sigma^+ \quad,\quad
e^{-\lambda\tilde\sigma^-} = e^{-\lambda\sigma^-}-\lambda\Delta
\quad.\label{214}\ee
The first relation in (\ref{214}) is consistent with the choice (\ref{211}),
since it implies that for $x^+ > x_0^+$ there is no incoming flux at the past 
null-infinity, i.e.
$$ \vev{T_{\tilde\sigma^+\tilde\sigma^+}}=0 \quad.\label{216}$$
The second relation in (\ref{214})
implies that the Hawking flux is the same as the initial thermal flux, i.e.
\be
\vev{T_{\tilde\sigma^-\tilde\sigma^-}}={N\lambda^2\over48\pi}\>.\label{218}
\ee
It is not difficult to see that, due to the Hawking radiation, 
the apparent horizon 
shrinks and meets the curvature singularity in a finite proper time.
The evaporating solution can be continuously matched to a static solution
on the null line $x^-=x^-_{int}$.
This solution coincides with the remnant geometry of \cite{BPP}, which
appears in the evaporation
process initiated by a gravitational collapse.
Therefore this 2d example confirms the intuition 
that the basic features of the evaporation process do not depend
on the way how the black hole was created.

\section{One-loop analytic model for Schwarzschild black hole evaporation}

Now we apply the idea of thermal bath removal to the model (\ref{14}).
The main problem which appears when trying to apply the 
RPS operator formalism to
the theory (\ref{14}) is that, in contrast to the theory (\ref{13}),
we do not know the general classical solution
for an arbitrary matter energy-momentum tensor $T_{\mu\nu}$.
However, if we want to describe the evaporation process of a black hole
which is initially in thermal equilibrium and then
the incoming thermal flux is switched off, the problem becomes simpler.

We start from the Vaidya solution (\ref{15}), and
in analogy with the 2d case, we take the following state $\ket{\Psi_0}$
\be \ket{\Psi_0} = \theta (v_0 - v ) \ket{0_V} \otimes \ket{0_U } + 
\theta (v - v_0 )\ket{0_v } \otimes \ket{0_U } \, , \label{31} \ee
where $V = 4M\exp(v/4M)$ and $U = -4M\exp(-u/4M)$ are the
Kruskal coordinates of the initial Schwarzschild black hole
and $M$ is its mass. Consistency
then requires that for $v < v_0$
\be \vev{ T_{\mu\nu} } = 0 \quad, \label{32}\ee 
which is satisfied if $T_{\mu\nu}$ is normal ordered with respect to the
Hartle-Hawking vacuum  $\ket{0_V } \otimes \ket{0_U }$. The incoming
and the outgoing flux are constant for $v < v_0$, and take the value
corresponding to the temperature $T = (8\pi M)\sp{-1}$
\be  \vev{ \tilde{T}_{vv} } =
\vev{ \tilde{T}_{uu} } = {N\over 48\pi (4M)\sp 2} \quad, \label{33}\ee 
where $\tilde{T}_{\mu\nu}$ denotes the operator obtained by normal ordering
${T}_{\mu\nu}$ with respect to the 
asymptotically flat coordinates $(u,v)$. For $v> v_0$, one obtains
\be \vev{T_{vv} } = - {N\over 48\pi (4M)\sp 2} = -\beta /2 \quad,\quad
\vev{ T_{uu} } = 0 \quad, \label{34}\ee 
where the first equation follows from
\bea
\vev{T_{vv}} &=& \bra{0_v} :T_{vv}:_V \ket{0_v} =
{\left({dV\over dv}\right)}\sp 2
 \bra{0_v} :T_{VV}:_V \ket{ 0_v} \nonumber \\
  &=& -{N\over 24\pi}{\left({dV\over dv}\right)}\sp 2 D_V (v) 
\quad,\nonumber\\ 
\eea 
and $D_V (v) = {v'''\over v'} - \frac32 {\left({v''\over v'}\right)}\sp 2$ 
is the Schwartzian derivative. The effective metric is then obtained by taking
the expectation value of the expression (\ref{15}), so that
\be
\vev{ds^2} =-\left( 1 + {1\over r}[\beta v \theta (v) - r_s ] \right) 
dv^2 + 2dvdr \>,\label{36}
\ee
where we have used (\ref{32}), (\ref{34}) and (\ref{16}). $r_s = 2M$ is the
Schwarzschild radius and we have set $v_0 = 0$. 
The effective metric (\ref{36}) is of the one-loop
order since it is only a function of $\vev{T_{\mu\nu}}$ and it does not depend
on $\vev{T_{\mu\nu}T_{\rho\sigma}}$ or on the higher-order energy-momentum
tensors correlation functions. 

For $v > 0$ the metric (\ref{36}) represents an
evaporating black hole whose mass is linearly decreasing with time. Such
a metric was previously studied in \cite{Hiscock}, where it was
ad hoc postulated and 
used to describe the evaporation phase of a black hole which was created
from a vacuum. Consequently, a flat spacetime was chosen for $v < 0$,  
instead of the Schwarzschild spacetime. The advantage
of our approach is that the operator formalism provides
metrics which are consistent with the boundary conditions. In this way one
avoids inconsistencies which may appear due to the ad hoc nature
of the procedure used in \cite{Hiscock}. For example, a flat
metric was chosen for $v> r_s/\beta$ in \cite{Hiscock}, and since the 
Hawking radiation is produced, one obtains a flat spacetime with non-zero
energy-momentum tensor. 
  
The line element (\ref{36}) can be written in the conformal
form (we will omit the expectation value)
\be
ds^2= -\theta (-v)\left(1-{r_s\over r}\right)dvdu + \theta (v)
{r\over z^2r_s}\left(1+z-2\beta z^2\right)dvd\tilde u \>, \label{37}
\ee
where 
\be
z={r\over \beta v - r_s} \>,\label{38}
\ee
and the coordinate $\tilde u$ is determined from the equation
\be
|1-\beta v / r_s |^{1/\beta} e^{\tilde u / r_s} =
|z - z_- |^{-A_-} |1- z/ z_{+}|^{-A_+} \quad.\label{39}
\ee
The constants $z_\pm$, $A_\pm$ are given by
\be
z_{\pm}= {1 \pm \sqrt{ 1 + 8\beta}\over 4\beta} \quad,\quad
A_{\pm} = {1 \over 2\beta}\left( 1 \pm {1\over\sqrt{ 1 + 8\beta}}
\right)\quad.\label{311}
\ee
Note that (\ref{39}) can be also viewed as the equation determining 
$r = r (\tilde u ,v)$.

The requirement that the conformal factor in (\ref{37}) is continuous at 
$v = 0$ gives 
\be
{\tilde u\over r_s}=-A_- \log\left| {r\over r_s} + z_- \right| - 
A_+ \log\left| 1+{r\over r_s z_+} \right| \>,\label{313}
\ee
where
\be
r + r_s \log\left| {r\over r_s}-1 \right| = -{u\over2} \>.\label{314}
\ee
The relations (\ref{313}) and (\ref{314})
determine the function $u = u (\tilde u )$.
One can now check that 
the incoming thermal flux has been removed for $v>0$, since $v$
remains the asymptotically flat coordinate
at the past null infinity $ \tilde u\rightarrow -\infty$. 

The scalar curvature of the effective metric is given by
\be
R={2\over r^3}\left(r_s -\beta v\theta (v) \right) \>,\label{315}
\ee
so that the curvature  singularity is at $r=0$.
The apparent horizon curve is determined by
$\partial_{v}r=0$, which gives
\be
r_{AH}=r_s - \beta v\theta (v)  \>. \label{316}
\ee
$r_{AH}$ decreases as the black hole evaporates, and the curve (\ref{316}) 
intersects the $r=0$ curve at
\be
v_{int}=r_s / \beta \quad,\quad {\tilde u}_{int} = \infty \quad,\label{317}
\ee
so that
for $v > v_{int}$ a naked singularity appears, see Fig. 1. 

Since the metric
(\ref{36}) is a one-loop approximation, it is valid only in the region
where $R l\sb P\sp 2 < 1$ (since the loop-expansion of the effective metric
is in 
$\vev{T\sp n}$, which is of the order of $(Rl\sb P\sp 2 )\sp n$). This gives 
that $r_{AH} > {\sqrt 2}l_P$,
which is the expected Planck length cutoff.
Note that if one defines a dynamical black hole mass $M_E$ as
$M_E = {1\over2}r_{AH}$, then
\be {dM_E\over dv} = -\beta / 2 \quad,\label{318}\ee 
which does not correspond to a thermal evaporation mass equation which is 
given by ${dM_E\over dv}\propto - M_{E}^{-2}$.
However, because $\beta$ is very small (in physical units it is given by
$ \beta = {N\over 384\pi}(m_P / M )\sp 2$ where $m_P$ is the Planck mass), 
the difference between
the non-thermal evaporation (\ref{318}) and a thermal one will be noticed
only when $r_{AH} < r_c \approx \frac45 r_s $. Therefore the evaporation is 
thermal for  $r_{AH} > r_c >> l_P $.

The Hawking flux ${\cal T}_{H}$ can be calculated from the expression
\be
 {\cal T}_{H}= \bra{\Psi_0 } T_{u_F u_F} \ket{\Psi_0 }
= \bra{0_U } T_{u_F u_F} \ket{0_U } =
-{N\over24\pi}D_{u_F}\left(U\right)\>,\label{319}
\ee
where $u_F$ is the asymptotically flat coordinate at the future null infinity
($v = \infty $). By using (\ref{39}) one can show that as
$v\rightarrow \infty $
\be
ds^2 \approx -{2z_+\over A_+}\left|z_+-z_-\right|^{-A_- / A_+ }
\left({\beta v\over r_s}\right)^{1- 1/ \beta A_+ }
e^{- \tilde u / r_s A_+ }dvd\tilde u \>.\label{320}
\ee
Hence the asymptotically flat coordinate $u_F$ is given by 
\be
 u_F = -A_+ r_s \left(e^{- \tilde u / r_s A_+ }-1\right)\>.\label{321}
 \ee
By using (\ref{39}) and the implicit relation
$\tilde u=\tilde u (u)$ defined by (\ref{313}) and (\ref{314}) one can 
work out the Hawking flux from (\ref{319})
\bea
{\cal T}_H &=& {\cal T}_0 |x + z_- |^{- 2 A_- / A_+ }
|1 + x / z_+ |^{-2}  {\Big[} 1 + 4\beta (x + 1 + x\sp{-1} ) \nonumber\\
&+& 4\beta\sp 2 (x\sp 2 + 2x + 4 ) -  {4\over A_+^2} {\Big ]}
\quad,\nonumber\\ 
\eea
where $x=r(\tilde u , v=0)/r_s = r(u , v=0) /r_s $. The behavior
of the flux is plotted in Fig. 2. 

For early times 
($\tilde u \rightarrow -\infty$ or $x \rightarrow \infty$) ${\cal T}_H$ is 
close to the initial thermal value 
\be
{\cal T}_H \approx {\cal T}_0 x\sp{-4\beta}
\approx {\cal T}_0 \exp \left( 4\beta \log (2\beta) +
4\beta\sp 2 \tilde u /r_s \right) \quad,\label{322a}
\ee
while for the late times ($\tilde u \rightarrow \infty$ or 
$x \rightarrow -z_-$)
it diverges at the 
end-point $x=-z_- = 1 - 2\beta + O (\beta^2 )$ as
\be
{\cal T}_H \approx {\cal T}_0 |x+z_- |^{-4\beta} (1 + 8\beta ) \quad.
\label{323}
\ee
This is an expected behaviour because of the presence of the naked singularity
at the end-point, and it is related to the fact that 
${dM_E \over dv}\ne 0$ at the end-point \cite{Hiscock}. 
However, as  the end-point is approached, the higher-order
loop corrections become relevant, and the one-loop approximation 
is expected to break down,
so that the one-loop divergence could be removed by the higher-loop
corrections. This actually happens in the CGHS case when the
two loop corrections are taken into account \cite{MR}. Therefore one can
expect that the higher-loop corrections will remove the naked singularity.

\section{Conclusions}

Note that our metric is a self-consistent semiclassical solution in the sense 
that its Einstein tensor is proportional to $\vev{T\sb{\mu\nu}}$ by 
construction. However, our metric does not
satisfy an additional requirement that the Hawking flux is finite at the future
null infinity \cite{Hiscock}. 
In our case this means that the higher-order quantum
corrections become important near the end-point. 
In the 2d case, the one-loop metric 
of \cite{BPP} satisfies the both criteria; however, it is defined only in the
weak-coupling region of the space-time, i.e. in the region where the 
higher-order quantum corrections can be neglected. Note that in our case the
flux stays very close to the thermal classical value until very late times
$\tilde u$. Therefore one could employ the BPP strategy of removing
the naked singularity by imposing a
strong-coupling boundary at $\tilde U ={\tilde U}_b \approx 0$,
where $ \tilde U = - 4M \exp(-\tilde u/4M)$, and then in the region
$\tilde U > {\tilde U}_b$, $V > V_{int}$ try to impose a static metric such that
it coincides with (\ref{37}) at $\tilde U = {\tilde U}_b$ 
for $V > V_{int} $. Also note that ${\tilde U} = 0$ line is tangential to the
$r =0$ curve at $V = V_{int}$, so that one has the same situation as in the BPP 
case. The diference now is that the value of the Hawking flux is infinite
at $\tilde U =0$, which is problematic. This is avoided
by putting the strong-coupling boundary at $\tilde U ={\tilde U}_b < 0$.

When $\tilde u = \tilde u_b$, then
$$ds^2 = - C_b (v) dvd\tilde u \quad, $$  
where 
$C_b (v) = 2 \left({\partial r\over \partial \tilde u}\right)|\sb{\tilde u_b}$,
so that a trivial solution in the region 
$\tilde U > {\tilde U}_b$, $V > V_{int} $ is 
$$ds\sp 2 = -d{\tilde u} d{\tilde v} \quad,$$
where $d{\tilde v} = C_b (v) dv$.
However, this is not a good solution because $\tilde u \ne u$ and 
$\tilde v \ne v$, which means
that radiation is present in the flat spacetime
region $\tilde U > {\tilde U}\sb{b}$, $\tilde V > V_{int}$. This is no 
surprise, because one expects
that the end-state geometry cannot be a flat space, but it should
be an asymptotically flat quantum corrected vacuum geometry, and the
corresponding one-loop effective action must contain additional counterterms
to the Polyakov-Liouville counterterm,
in analogy to the BPP case \cite{BPP}. This quantum vacuum geometry cannot
be determined within our construction. However, it is clear how our
construction can be
extended. One should find a more general class of classical solutions than the
Vaidya solutions, and then quantize them according to our approach. These
classical solutions could be obtained either approximately or by using the
global symmetries of the theory \cite{CNT}. 

\bigskip

\noindent {\bf Acknowledgements:} A.M. would like to thank Generalitat 
Valenciana for the financial support.

\newpage

\begin{figure}[h]
\centerline{\psfig{figure=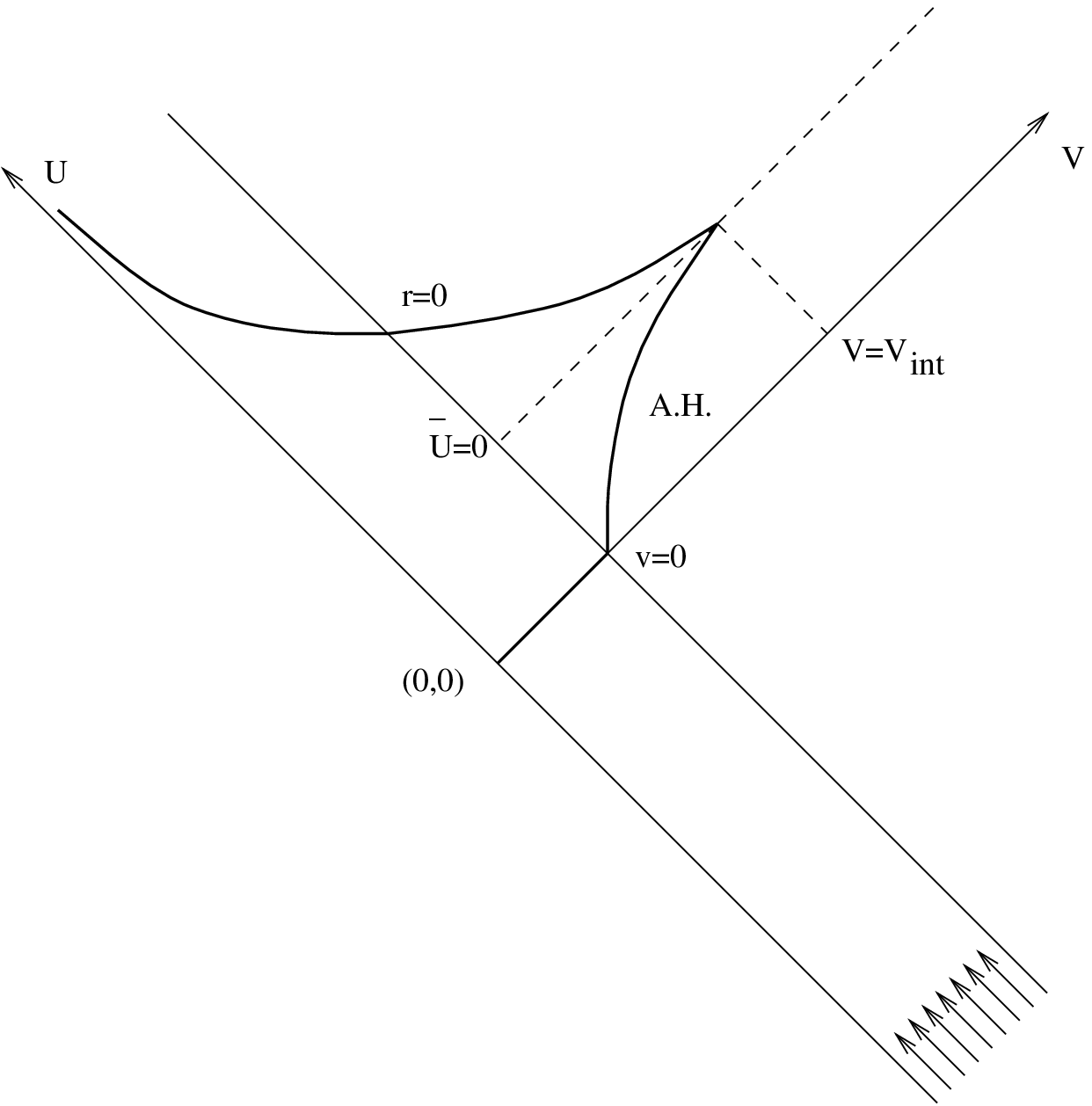,width=10cm}}
\caption{Kruskal diagram of the one-loop geometry.}
\end{figure}

\newpage

\begin{figure}[h]
\centerline{\psfig{figure=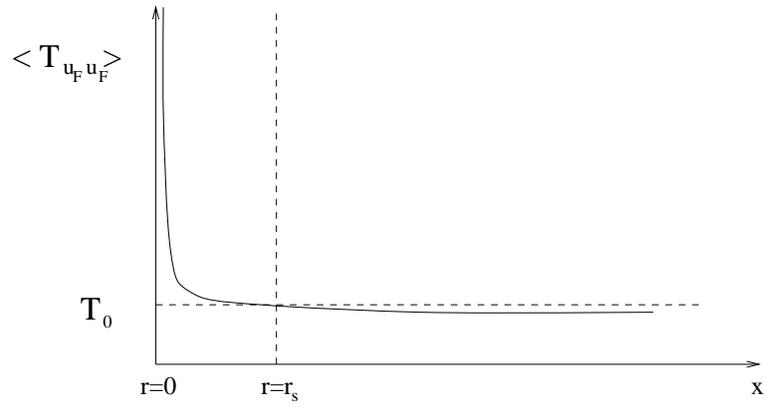,width=10cm}}
\caption{One-loop Hawking flux.}
\end{figure}

\end{document}